\documentstyle[sprocl]{article}

\input{psfig.sty}

\def\drop#1{}

\newcommand{\equ}[1]{~Eq.~(\ref{#1})}

\newcommand{\vev}[1]{\left\langle #1 \right\rangle}

\newcommand{\sla}{\raise.15ex\hbox{$/$}\kern -.8em}

\newcommand{\half}{\frac 1 2}

\newcommand{\cb}{{\bar c}}

\renewcommand{\L}{\Lambda}
\newcommand{\LQCD}{\L_{\overline{MS}}}

\newcommand{\m}{\mu}\newcommand{\n}{\nu}
\newcommand{\mn}{{\mu\nu}}
\newcommand{\eps}{\varepsilon}
\newcommand{\bra}{\langle}\newcommand{\ket}{\rangle}
\newcommand{\be}[1]{\begin{equation}\label{#1}}
\newcommand{\ee}{\end{equation}}
\newcommand{\ba}[1]{\begin{eqnarray}\label{#1}}
\newcommand{\ea}{\end{eqnarray}}

\title{Mass Generation, Ghost Condensation and Broken Symmetry:
$SU(2)$ in Covariant Abelian Gauges}
\author{Martin Schaden$^\dagger$}
\address{New York University, Physics Department,
 4 Washington Place, NY 10003}
\date{\today}
\begin{document}
\maketitle
\begin{abstract}
The local action of an $SU(2)$ gauge theory in general covariant
Abelian gauges and the associated equivariant BRST symmetry that
guarantees the perturbative renormalizability  of the model are
given. A global SL(2,R) symmetry of the model is spontaneously
broken by ghost-antighost condensation at arbitrarily small
coupling. This leads to propagators that are finite at Euclidean
momenta for all elementary fields except the Abelian ``photon''.
Ward Identities show that the symmetry breaking gives rise to
massless BRST-quartets with ghost numbers $(1,2,-2,-1)$ and
$(0,1,-1,0)$. The latter quartet is interpreted as due to an
Abelian Higgs mechanism in the dual description of the model.
\end{abstract}
\footnotetext{$^\dagger$ Present address: 144 Broadway, Ocean
Grove, NJ 07756; email: m.schaden@att.net}

\section{Introduction}
Although QCD is asymptotically free, the perturbative analysis of
the high-temperature phase is plagued by infrared (IR)
divergences\cite{Ka89}. The best one can presently achieve
perturbatively at high temperatures is a resummation of the
infrared-safe contributions\cite{Pi89}. The situation is somewhat
embarrassing, since one naively might expect that the high
temperature phase of an asymptotically free theory is
perturbative. The IR-problem encountered in the perturbative high
temperature expansion is part of the more general problem of
defining a non-Abelian gauge-fixed theory on a compact Euclidean
space-time without boundaries, such as a hypertorus. It was
shown\cite{Ba98} that normalizable ghost zero-modes in this case
cause the partition function to vanish in conventional covariant
gauges. An equivariant BRST construction was used to eliminate
these ghost zero-modes associated with global gauge invariance at
the expense of a non-local quartic ghost interaction. This
interaction leads to ghost-antighost condensation at arbitrarily
small coupling\cite{Ba98,Sc98}.

The equivariant gauge-fixing procedure later was used to reduce
the structure group of an $SU(2)$ lattice gauge theory (LGT) to a
physically equivalent Abelian LGT with a $U(1)$ structure
group\cite{Sc99}. This lattice formulation is the only known
definition of an non-Abelian (equivariant) BRST symmetry that is
not just valid perturbatively. The associated local quartic ghost
interaction leads to ghost-antighost condensation in this case as
well. The starting point of this investigation is a transcription
of the partially gauge-fixed SU(2)-LGT to the continuum using the
equivariant BRST algebra. Note that the global equivariant
BRST-symmetry makes the construction of the critical continuum
model of the partially gauge fixed non-Abelian LGT practically
unique. This aspect of the equivariant BRST symmetry is of some
interest in itself\cite{Pa00}.

\section{The Model}
In a Euclidean space-time, the critical continuum action of the
lattice model\cite{Sc99} is uniquely specified by the BRST
algebra, the field content and power counting. Decomposing the
non-Abelian SU(2) connection $\vec A_\mu=(W_\mu^1,W_\mu^2, A_\mu)$
in terms of two real vector-bosons (or one complex one) and a
U(1)-connection $A_\mu$ (the ``photon'' of the model), the loop
expansion is defined by the Lagrangian,
\be{L}
{\cal L}={\cal L}_{\rm inv.}+{\cal L}_{\rm AG}+{\cal L}_{\rm aGF}\ .
\ee
Here ${\cal L}_{\rm inv.}$ is the usual $SU(2)$-invariant Lagrangian
written in terms of the vector bosons and the
photon\footnote{Latin
indices take values in $\{1,2\}$ only, Einstein's summation
convention applies and $\eps^{12}=-\eps^{21}=1$, vanishing
otherwise. All results are in the $\overline{MS}$ renormalization
scheme.},
\be{Linv}
{\cal L}_{\rm inv.}={\cal L}_{\rm matter}+ {\frac 1 4} (G_\mn G_\mn+ G_\mn^a
G_\mn^a)\ ,
\ee
with
\ba{G}
G_\mn&=&\partial_\m A_\n-\partial_\n A_\m -g\eps^{ab} W_\m^a W_\n^b\nonumber\\
G_\mn^a&=&D^{ab}_\m W_\n^b- D^{ab}_\n W_\m^b\nonumber\\
&=&\partial_\m W_\n^a
-\partial_\n W_\m^a+g\eps^{ab}(A_\m W_\n^b-A_\n W_\m^b)\ .
\ea
${\cal L}_{\rm AG}$ partially gauge-fixes to the maximal
Abelian subgroup $U(1)$ of $SU(2)$ in a covariant manner,
\be{MAG}
{\cal L}_{AG}=b^a F^a-\frac{\alpha}{2} b^a b^a-\cb^a M^{ab} c^b
-g^2 \frac{\alpha}{2} (\cb^a\eps^{ab} c^b)^2\ ,
\ee
with
\ba{defs}
F^a&=& D_\m^{ab} W_\m^b=\partial_\m W_\m^a+ g A_\m\eps^{ab} W_\m^b\nonumber\\
M^{ab}&=& D_\m^{ac} D_\m^{cb} + g^2\eps^{ac}\eps^{bd} W_\m^c
W_\m^d\ .
\ea
Note that $L_{U(1)}=L_{\rm inv.}+{\cal L}_{\rm AG}$ is invariant
under $U(1)$-gauge transformations {\it and} under an on-shell
BRST symmetry $s$ and anti-BRST symmetry $\bar s$, whose action on the
fields is
\begin{displaymath}
\begin{array}{rclcrcl}
s A_\m&=&g\eps^{ab}c^aW_\m^b&& \bar sA_\m&=&g\eps^{ab}\cb^aW_\m^b\\
s W_\m^a&=& D_\m^{ab} c^b&& \bar s W_\m^a&=& D_\m^{ab} \cb^b\\
s c^a&=&0&&\bar s \cb^a&=&0\\
s \cb^a&=&b^a&&\bar s c^a&=&-b^a\\
s b^a&=&s^2\cb^a={\frac g 2}\eps^{ab}\cb^b c^c\eps^{cd}c^d &&\bar
s b^a&=&-{\frac g 2}\eps^{ab}c^b \cb^c\eps^{cd}\cb^d \ ,
\end{array}
\end{displaymath}
\nopagebreak\vspace{-1.0cm}\be{brs}\ee
with an obvious extension
to include matter fields. Contrary to most other proposals for
mass generation\cite{Cu76,Bl96} the BRST algebra\equ{brs} closes
on-shell on the set of $U(1)$-invariant functionals: on
functionals that depend only on $W,A,c$ and the matter fields,
$s^2$ for instance effects an infinitesimal U(1)-transformation
with the parameter ${\frac g 2}\eps^{ab}c^a c^b$. The algebra
\equ{brs} defines an equivariant cohomology. The lattice
regularization of this symmetry\cite{Sc99} ensures that the model
is renormalizable and unitary. Perturbative renormalizability and
unitarity has recently also been algebraically established for
this model\cite{Fa01}. Note that the physical sector comprises
states created by composite operators of $A,W$ and the matter
fields in the equivariant cohomology of $s$ (or $\bar s$). They
are BRST closed {\it and} $U(1)$-invariant.

Expectation values of physical observables of the U(1)-LGT are the
same\cite{Sc99} as those of the original $SU(2)$-LGT for any
$\alpha>0$. Note that setting $\alpha=0$ and formally solving the
constraint\cite{Re98} $F^a=0$ is {\it not} the same as taking the
limit $\alpha\rightarrow 0$. The non-perturbative reason is
exhibited by the lattice calculation\cite{Sc99}: without the
quartic ghost interaction, Gribov copies of a configuration
conspire to give {\it vanishing} expectation values for all
physical observables. No matter how small, the quartic ghost
interaction is necessary for a normalizable partition function and
expectation values of physical observables that are identical with
those of the original SU(2)-LGT. Without Abelian ghosts, a quartic
ghost interaction is generated by perturbative corrections even in
the $\alpha\rightarrow 0$ limit\cite{Mi85}. Currently, a lattice
regularization only exists for the continuum model described
by\equ{L} -- with decoupled Abelian ghosts and an SL(2,R)
symmetry.

In the continuum model ${\cal L}_{\rm aGF}$ in\equ{L} can be added
``by hand'' to fix the remaining $U(1)$ gauge invariance and
define the perturbative series of the continuum model
unambiguously. Note that this Abelian gauge fixing does not
introduce new ghosts. I will assume a conventional covariant
gauge-fixing term,
\be{aGF}
L_{\rm aGF}=\frac{(\partial_\m A_\m)^2}{2\xi}\ .
\ee
None of the following conclusions  depend on the gauge-fixing of
the Abelian subgroup -- they in particular do not depend on $\xi$.

\section{Ghost Condensation and the Spontaneously Broken SL(2,R) Symmetry}
The Lagrangian \equ{L} is invariant under a global bosonic SL(2,R)
symmetry generated by
\be{SL2}
\Pi^+=\int c^a(x)\frac{\delta}{\delta \cb^a(x)}\ \ ,\ \
\Pi^-=\int \cb^a(x)\frac{\delta}{\delta c^a(x)}\ ,
\ee
and the ghost number $\Pi=[\Pi^+,\Pi^-]$. This SL(2,R) symmetry is
inherited from the lattice regularized model\cite{Sc99} and
therefore is not anomalous. The conserved currents corresponding
to $\Pi^\pm$ are U(1)-invariant and BRST, respectively anti-BRST
exact,
\be{currents}
j^+_\m=c^a D^{ab}_\m c^b=s c^aW^a_\m\ ,\ \ j_\m^-=\cb^a
D^{ab}_\m\cb^b=\bar s \cb^aW_\m^a\ .
\ee
Because the currents\equ{currents} are (anti)-BRST exact, a
spontaneously broken SL(2,R) symmetry is accompanied by a
BRST-quartet of massless Goldstone states with ghost numbers
$1,2,-2$ and $-1$. They are U(1)-invariant $c-W$,$c-c$, $\cb-\cb$
and  $\cb-W$ bound states. Such BRST quartets do not contribute to
physical quantities\cite{Ku78} like the free energy\footnote{This
is analogous to the decoupling of the BRST quartet of the weak
interaction in $R_\xi$ gauges\cite{Ku78}.}. The spontaneous
symmetry breaking in this sense is similar to a (dynamical) Higgs
mechanism in the adjoint.

An order parameter for the spontaneous breaking of the
SL(2,R) symmetry is
\be{opar}
\bra\cb^a\eps^{ab}c^b\ket=\half\bra\Pi^-
(c^a\eps^{ab}c^b)\ket=-\half\bra\Pi^+ (\cb^a\eps^{ab} \cb^b)\ket\
.
\ee
One can argue for such a ghost condensate in the background of a
degenerate orbit such as that of a non-Abelian monopole. The
solution to $F^a=0$ in this case is not unique locally and the
operator $M^{ab}$ in such a background has an even number of ghost
zero-modes that interact via the quartic ghost interaction only.
Similar to states at the Fermi surface of a BCS-superconductor,
condensation occurs in the attractive channel of the quartic
interaction. The notion that the predominance of degenerate orbits
may trigger ghost condensation can be sharpened considerably by
considering the Ward identities associated with the broken SL(2,R)
symmetry.

Using that the currents\equ{currents} of the broken $SL(2,R)$
symmetry are U(1)-invariant and (anti)-BRST exact, one can show
that
\be{pole2}
\vev{s(\cb^a\eps^{ab}\cb^b)(0)\quad c^a
W_\m^a(x)}=\frac{\vev{\cb^a\eps^{ab}\cb^b}x_\m}{\pi^2
x^4}=-\vev{\bar s(c^a\eps^{ab}c^b)(0)\quad \cb^a W_\m^a(x)}\ ,
\ee
implying massless asymptotic states with ghost numbers $\pm1$.
Together with the Goldstone states with ghost number $\pm2$ these
form the (anti)-BRST-quartets of the spontaneously broken SL(2,R)
symmetry.

However, the symmetries imply even more massless asymptotic states
when the SL(2,R) symmetry is spontaneously broken. Since,
\ba{equiv1}
s(\cb^a\eps^{ab}\cb^b)=& 2 b^a\eps^{ab}\cb^b & =-2\bar
s(c^a\eps^{ab}\cb^b)\nonumber\\
\bar s(c^a\eps^{ab}c^b) =& -2 b^a\eps^{ab} c^b  &=-2
s(\cb^a\eps^{ab}c^b)\ ,
\ea
the spontaneous symmetry breaking is associated with yet another
(anti)-BRST-quartet of massless states whose ghost numbers are
$0,1,-1,0$. Using\equ{equiv1} in\equ{pole2} gives
\be{Hgold1} -\vev{\bar s (c^a\eps^{ab}\cb^b(0))\quad c^a
W_\m^a(x)}=\frac{\vev{\cb^a\eps^{ab}\cb^b}x_\m}{2\pi^2 x^4}=\vev{
s (\cb^a\eps^{ab}c^b(0))\quad \cb^a W_\m^a(x)}.
\ee
There are thus asymptotic massless states with vanishing ghost
number
\be{Hgold0}
\vev{\cb^a\eps^{ab}c^b(0)\quad
\bar s( c^a W_\m^a(x))}=\frac{\vev{\cb^a\eps^{ab}c^b}
x_\m}{2\pi^2 x^4} = -\vev{\cb^a\eps^{ab}c^b(0)\quad s( \cb^a
W_\m^a(x))}\ .
\ee
\equ{Hgold0} is consistent with the absence of a massless pole in
correlations of the ghost number current
\be{ghostnr}
j^0_\m=-\half [s(\cb^a W_\m^a)+\bar s(c^a W_\m^a)]=\half[\cb^a
D^{ab} c^b+ c^a D^{ab} \cb^b]\ .
\ee

The BRST-quartet of massless states implied by\equ{Hgold0} could
also be the result of a spontaneously broken Abelian gauge
symmetry. To see this, consider a U(1) gauge model with a
self-interacting charged scalar $\Phi(x)$ that transforms as
$\delta \Phi(x)= i\Phi(x)\delta\theta(x)$, the Higgs mechanism
leads to the relation
\be{Higgs}
\vev{\Phi(0) \ \ \frac{\delta
S}{\delta\theta(x)}}=i\delta^4(x)\vev{\Phi}\ ,
\ee
where $S$ is the gauge-fixed action of this Abelian model.  In
covariant gauges that do not break the global U(1) invariance,
$\delta S/\delta\theta(x)=-i\partial_\m J_\m(x)$ is a BRST-exact
current $J_\m(x)=s \bar C_\m(x)$. The fundamental
BRST-quartet\cite{Ku78} remains massless in gauges that do not
break the global U(1) invariance under consideration.

If we identify $\cb^a W_\m^a$ with the current $\bar C_\m$ and
$\cb^a\eps^{ab}c^b$ with the charged Higgs scalar $\Phi(x)$ of the
effective Abelian Higgs model, the right hand side of\equ{Hgold0}
is the analogue of\equ{Higgs}.

Note that the spontaneously broken Abelian (gauge) symmetry in the
present case has vanishing ghost number but cannot be the
chromo-electric U(1) nor the ghost number of the covariant MAG:
the scalar $\cb^a\eps^{ab} c^b$ is neutral under both.
\equ{Hgold0} implies that a massless asymptotic state with
vanishing chromo-electric charge and ghost number is created by
the longitudinal part of the composite vector field
\be{B}
B_\m:=b^a W_\m^a-\partial_\m(\cb^a c^a)=\half [s(\cb^a
W_\m^a)-\bar s(c^a W_\m^a)]\ .
\ee
The fact that $B_\m$ does not mix with $A_\m$ under
renormalization supports the conjecture that the chromo-electric
U(1) symmetry of MAG remains unbroken by ghost condensation in
this channel. It is tempting to assume that the transverse part of
$B_\m$ has non-vanishing overlap with the {\it dual} photon of the
model. Since the vector bosons are expected to be massive in the
confining phase, this is not in contradiction with a dual Higgs
mechanism\cite{tH81}.

The dynamical formation of a
\be{kondo}
\vev{W^a_\m W^a_\m-\alpha \cb^a c^a}
\ee
 condensate is discussed by Kondo\cite{Ko01}. It generates effective
masses for the vector bosons and ghosts without apparently
breaking any continuous symmetries of the model. Contrary to the
condensate of\equ{opar}, the condensate of\equ{kondo} is not
easily linked to a dual Abelian Higgs mechanism and leads to power
corrections of dimension 2 in physical correlation functions. [By
contrast, the leading power corrections due to the condensate
of\equ{opar} in physical correlation functions are of dimension
4.] The condensate of\equ{kondo} therefore could be characteristic
for the Coulomb phase\cite{tH81} of the model. This interesting
possibility will not be further pursued here.

\section{Perturbations in the Broken Phase}
To investigate the perturbative consequences of
$\bra\cb^a\eps^{ab}c^b\ket\neq 0$, the quartic ghost interaction
in\equ{MAG} is linearized with the help of an auxiliary scalar
field $\rho(x)$ of canonical dimension two. Adding the quadratic
term
\be{Stratanovic}
{\cal L}_{\rm aux}=\frac{1}{2 g^2\alpha} (\rho/\sqrt{Z}
-g^2\alpha\sqrt{Z}\cb^a\eps^{ab} c^b)^2
\ee
to the Lagrangian of\equ{L}, the tree level quartic ghost interaction
 vanishes at $Z=1$
and is then formally of $O(g^4)$, proportional to $g^2$ and
$Z-1$\footnote{The discrete symmetry $ c^a\rightarrow \cb^a,\
\cb^a\rightarrow -c^a,\ \rho\rightarrow -\rho$ relating $s$ and
$\bar s$ also ensures that $\rho$ only mixes with
$\cb^a\eps^{ab}c^b$.}.

The perturbative expansion about a {\it non-trivial} solution
$\langle\rho\rangle=v\neq 0$ to the gap equation
\be{gap}
\frac{v}{g^2\alpha}=\left.\bra c^a(x)\eps^{ab}{\bar c}^b(x)\ket
\right|_{<\rho>=v}\ ,
\ee
turns out to be well behaved in the infrared. Note that \equ{gap}
is $U(1)$-invariant and therefore does not depend on the $U(1)$
gauge-fixing\equ{aGF}.  Let us for the moment assume that a unique
non-trivial solution to\equ{gap} exists in some gauge $\alpha$; we
return to this conjecture below.  The consequences for the
IR-behavior of the model are dramatic. Defining the quantum part
$\sigma(x)$ of the auxiliary scalar $\rho$ by
\be{decomp}
\rho(x)=v+\sqrt{\alpha}\sigma(x) \ \ {\rm with}\ \
\bra\sigma\ket=0\ ,
\ee
the momentum representation of the Euclidean ghost propagator at tree
level becomes
\be{ghost}
\bra c^a \cb^b\ket_p
=\frac{p^2\delta^{ab}+v\eps^{ab}}{p^4+v^2}\ .\\
\ee
Feynman's parameterization of this propagator allows an evaluation
of loop integrals using dimensional regularization  that is only
slightly more complicated than usual. More importantly, the ghost
propagator is regular at Euclidean momenta  when $v\neq 0$. Its
complex conjugate poles at $p^2=\pm i v$ can furthermore not be
interpreted as due to asymptotic ghost states\cite{St99}.

When $v\neq 0$, the $W$-boson is massless only at tree level and
(see Fig.~1) acquires the {\it finite} mass $m_W^2=g^2
|v|/(16\pi)$ at one loop,

\bigskip\medskip
\vbox{
\be{Wmass}
\hskip 4truecm {=\frac{g^2|v|}{16\pi}\delta_\mn \delta^{ab}\ .}
\ee
\vskip-1.5truecm \hskip
.5truecm\psfig{figure=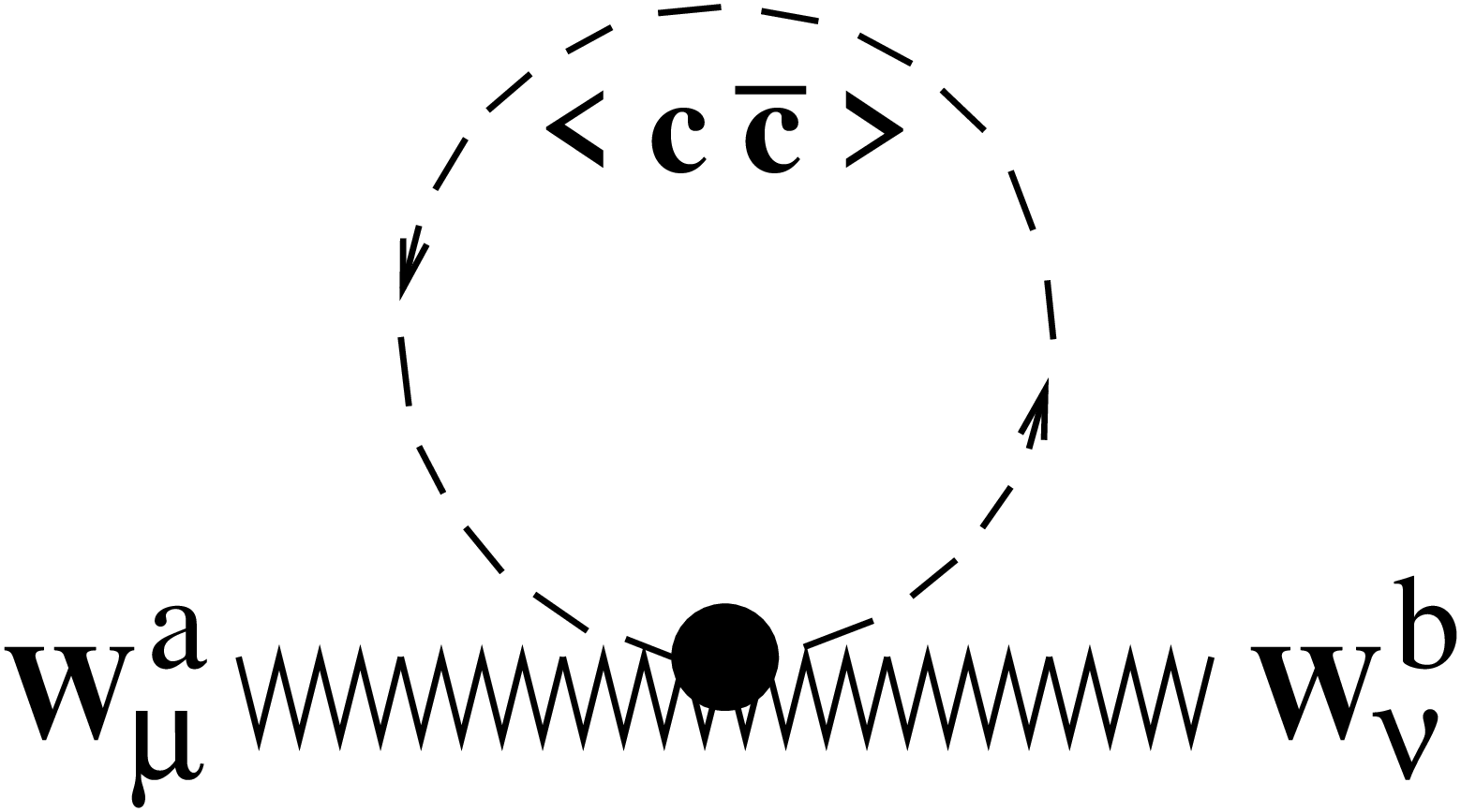,height=2.0truecm}
\\*
{\small\noindent Fig.~1. The finite one-loop contribution to the
$W$ mass.}} \vskip 5pt Technically, the one-loop contribution is
finite because the integral in\equ{Wmass} involves only the
$\delta^{ab}$-part of the ghost propagator\equ{ghost}. Since
$p^2/(p^4+v^2)=- v^2/(p^2(p^4+ v^2))+1/p^2$, the $v$-dependence of
the loop integral is IR- {\it and} UV-finite. The quadratic
UV-divergence of the $1/p^2$ subtraction at $v=0$ is cancelled by
the other, $v$-independent, quadratically divergent one-loop
contributions -- (in dimensional regularization this
scale-invariant integral vanishes by itself). $m_W^2$ furthermore
is {\it positive} due to the overall minus sign of the {\it ghost}
loop. The sign of $m_W^2$ is crucial, for it indicates that the
model is {\it stable} and (as far as the loop expansion is
concerned) does not develop tachyonic poles at Euclidean $p^2$ for
$v\neq 0$. Conceptually, the local mass term proportional to
$\delta_\mn\delta^{ab}$ is finite due to the BRST
symmetry\equ{brs}, which excludes a mass counter-term. The latter
argument implies that contributions to  $m^2_W$ are finite to all
orders of the loop expansion.

\hskip 1.5truecm\psfig{figure=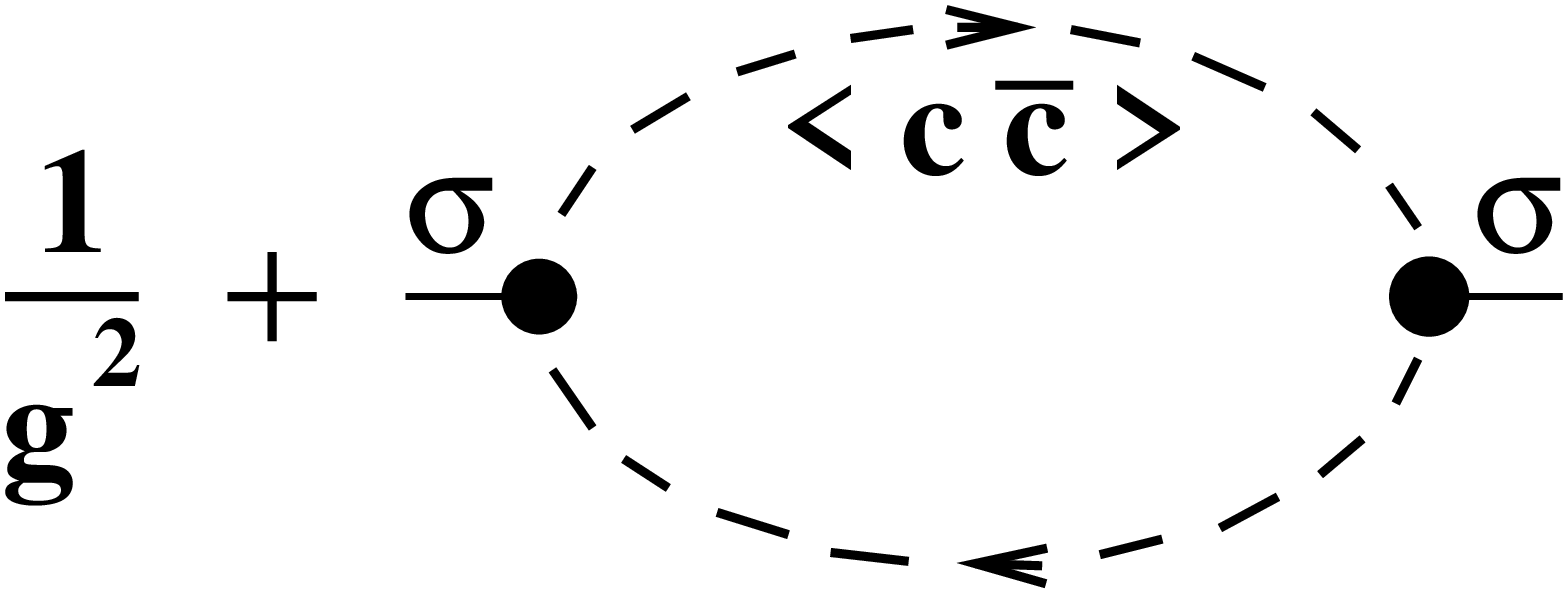,height=1.5truecm}
\\*
{\small\noindent Fig.~2. $\Gamma_{\sigma\sigma}(v,p^2)$ to
order $g^0$.}
\vskip 5pt
If the model is stable at $v\neq 0$, the 1PI 2-point
function $\Gamma_{\sigma\sigma}(v, p^2)$ of the scalar must not vanish
at Euclidean $p^2$ either. To order $g^0$, $\Gamma_{\sigma\sigma}(p^2)$ is
given by the $1/g^2$ term that arises from \equ{Stratanovic} upon
substitution of\equ{decomp} and the one-(ghost)-loop contribution
shown in Fig.2. Since a non-trivial solution to the gap
equation\equ{gap} relates $1/g^2$ to a loop integral of zeroth order
in the coupling, we may use
\equ{gap} to lowest order to  obtain a  ``tree-level''
expression for $\Gamma_{\sigma\sigma}(v,p^2)$ of order $g^0$.
Evaluating the loop integrals, one obtains the real, positive and
monotonic function
\ba{ss}
\Gamma_{\sigma\sigma}(x:=\frac{\sqrt{\alpha v^2}}{p^2})&=&\left\{\frac{
-1+2\sqrt{1-4ix}\,{\rm acoth}(\sqrt{1-4ix})}{32\pi^2\alpha^{-1}}\right\}\nonumber\\
&& + \left\{x\rightarrow -x\right\} \ .
\ea
%
$\Gamma_{\sigma\sigma}(p^2\geq 0)\ge\alpha/(16\pi^2)$ to order
$g^0$  establishes the perturbative stability of a non-trivial
solution to\equ{gap} and the fact that this solution is a minimum
of the one-loop effective potential.

An expansion about a  solution $v\neq 0$ to the gap equation thus
has lowest order propagators that are regular at Euclidean momenta
for all the elementary fields  except the photon $A_\m$ (if all
the matter fields are massive).  The polarization of the photon
vanishes at $p^2=0$ due to the $U(1)$-symmetry -- regardless of
the value of $v$.  Taking into account that the massless Goldstone
quartets associated with this symmetry breaking decouple from
physical quantities, the situation for $v\neq 0$ is thus rather
similar to QED with an unorthodox massive matter content
(extending the notion of ``massive matter'' to include ghosts and
other unphysical fields).

\section{The Gap}
To complete the argument, we solve\equ{gap} for small coupling. To
lowest order in the loop expansion, the relation between the
renormalized couplings $g,\alpha$, the renormalization point $\m$
and an expectation value $v\neq 0$ implied by \equ{gap} is
\be{solg}
\ln{\frac{v^2}{\m^4}}=-\frac{16\pi^2}{\alpha g^2}+2+O(g^2)\ .
\ee
The anomalous dimension $\gamma_v$ of the expectation value is
simultaneously found to be\footnote{This corrects the error
in\cite{Sc99a} of ignoring the corrections of order $g^2$ in
$Z=1+O(g^2)$}
\be{gammav}
\gamma_v=-\frac{d\ln Z_v}{d\ln \m}=\frac{g^2}{8\pi^2} (\alpha -3)
+O(g^4)\ .
\ee

Using the relation between $\m$, $g^2$ and the asymptotic scale
parameter $\LQCD$, we may rewrite \equ{solg} as
\be{solQCD}
\ln{\frac{v^2}{\LQCD^4}}=\frac{16\pi^2}{g^2}\left(\frac{2}{\beta_0}-
\frac{1}{\alpha}\right)+2+O(\ln{g}, g^2)\ ,
\ee
where $\beta_0$ is the lowest order coefficient of the
$\beta$-function of this model ($\beta_0=(22-2 n_f)/3$ with $n_f$
quark flavors in the fundamental representation as matter). Apart
from an anomalous dimension, the non-trivial solution $v$ at
sufficiently small coupling is thus proportional to the physical
scale $\LQCD^2$ in the {\it particular} gauge  $\alpha=\beta_0/2$.
The anomalous dimension $\gamma_v$ in\equ{gammav} furthermore is
of order $g^4$ at $\alpha=3$.  For $n_f=2$ quark flavors, the
terms of order $\ln{g}$ in\equ{solQCD} thus also vanish in the
particular gauge $\alpha=\beta_0/2=3$ and higher order corrections
to the asymptotic value of $v$ at small $g^2$ are analytic in
$g^2$. With $n_f=2$ flavors, one can expand the model about
\be{start}
v^2=e^2\LQCD^4(n_f=2) (1+O(g^2))
\ee
in the gauge $\alpha=\beta_0/2$, and determine the $O(g^2)$
corrections in \equ{start} order by order in the loop expansion of
the gap equation\equ{gap}. Note that this behavior is surprisingly
consistent with the previous observation\cite{Sc98} for $SU(n)$ in
generalized covariant gauges that the lowest order solution to the
gap equation remains accurate to order $g^2$ at any finite order
of the loop expansion in the gauge $\alpha=\beta_0/n$ when there
are $n_f=n$ light quark flavors. This {\it does not} mean that
other gauges are any less physical, but it does single out
$\alpha=\beta_0/n=3$ as a {\it critical} gauge in which the {\it
perturbative} evaluation of the gap equation\equ{gap} is
consistent for sufficiently small values of $g^2$. (In QED the
hydrogen spectrum to lowest order is most readily obtained in
Coulomb gauge, although it evidently does not depend on the chosen
gauge. In the present case asymptotic freedom determines an
optimal gauge for solving the gap equation at small coupling.)

At the one-loop level,  \equ{gap} has a unique non-trivial
solution in any gauge $\alpha\neq 0$ and $\Gamma_{\sigma\sigma}$
of\equ{ss} shows that it corresponds to a minimum of the one-loop
action. In the limit $\alpha\rightarrow 0$ at finite coupling, the
non-trivial one-loop solution \equ{solg} coincides with the
trivial one. On the other hand, some of the couplings in the
non-linear gauge-fixing ${\cal L}_{\rm AG}$ become large in this
limit, invalidating any perturbative analysis.

The highly singular behavior of the model when $\alpha\sim 0$ is
already apparent in the divergent part of the $W$ self-energy to
one loop. The corresponding anomalous dimensions $\gamma_W$ and
$\gamma_\alpha$ of the vector boson and the gauge parameter are
\ba{anom}
\gamma_W&=&-\frac{d\ln Z_W}{d\ln
\mu}=\frac{g^2}{8\pi^2}\left(\beta_0-\frac{9}{2}- \frac{\alpha}{2}-\xi
\right)+O(g^4)\ ,\nonumber\\
\gamma_\alpha &=&\frac{d\ln{\alpha}}{d\ln \m} =
-\frac{g^2}{8\pi^2}\left(\frac{3}{\alpha}+6-\beta_0 +
\alpha\right)+O(g^4).
\ea
Gauge dependent interaction terms proportional to
$g/\alpha$ at one loop thus lead to a term of order $g^2/\alpha^2$ in the
longitudinal part of the $W$ self-energy only.
The transverse part of the $W$~self-energy is
regular in the limit $\alpha\rightarrow 0$. Taking $\alpha$ to vanish
thus is rather tricky:
\equ{anom} implies that the longitudinal part of the $W$-propagator at
one loop is proportional to $3
g^2 p^2\ln(p^2)$ at large momenta and no longer vanishes in this limit.
Higher order loop corrections similarly
contribute to the longitudinal propagator as $\alpha\rightarrow 0$.
$\gamma_\alpha$ does not depend on the gauge parameter $\xi$ at one loop,
due to an Abelian Ward identity that also gives the QED-like
relation\cite{Re98} $Z_A=Z_g^{-2}=Z_\xi$  between the renormalization
constants of the photon,  of the coupling $g$
and of the gauge parameter $\xi$.

The anomalous dimension of the gauge
parameter $\alpha$ at sufficiently small $g^2$
is negative for positive values of $\alpha$ when
$\beta_0<6+2\sqrt{3}$. With $\gamma_\alpha<0$, the effective gauge
parameter tends to decrease at higher renormalization scales $\m$
and direct integration of \equ{anom} gives a vanishing $\alpha$ at
a {\it finite} value of the coupling $g^2$. As already noted
above, the loop expansion, however, is valid only if $g^2\ll 1$
and $g^2/\alpha\ll 1$. But \equ{anom} does show that there is no
{\it finite} UV fixed point for the gauge parameter and that
$\alpha$ effectively  vanishes at least as fast as $g^2$ as
$\m\rightarrow \infty$ for any fixed gauge at finite $g^2$.
\equ{start} nevertheless is the asymptotic solution to \equ{gap}
in the sense that it is  valid at arbitrary small coupling if the
gauge at that coupling is chosen to be $\alpha(g)=\beta_0/2$.

The existence of a (unique) non-trivial solution to the gap
equation can be viewed as a consequence of the scale
anomaly\cite{Sc98}. The renormalization point dependence
of\equ{solg} and the associated UV-divergence of the loop integral
are an indication of this.

I would like to thank D.~Kabat, D.~Zwanziger and R.~Alkofer for
suggestions, L.~Spruch for his continuing support, and L.~Baulieu
for encouragement and the organizers for this very stimulating
conference.

\end{document}